\documentclass[draft]{aipproc} 
\usepackage{amsfonts} 
\usepackage{epsf}

\def\llra{\relbar\joinrel\longrightarrow}              
\def\mapupdown#1#2{\smash{\mathop{\llra}\limits^{#1}_{#2}}} 

\catcode`\@=11

\def\Let@{\relax\iffalse{\fi\let\\=\cr\iffalse}\fi}
\def\vspace@{\def\vspace##1{\crcr\noalign{\vskip##1\relax}}}
\def\multilimits@{\bgroup\vspace@\Let@
 \baselineskip\fontdimen10 \scriptfont\tw@
 \advance\baselineskip\fontdimen12 \scriptfont\tw@
 \lineskip\thr@@\fontdimen8 \scriptfont\thr@@
 \lineskiplimit\lineskip
 \vbox\bgroup\ialign\bgroup\hfil$\m@th\scriptstyle{##}$\hfil\crcr}
\def\Sb{_\multilimits@}
\def\endSb{\crcr\egroup\egroup\egroup}
\def\Sp{^\multilimits@}

\layoutstyle{6x9}

\begin{document}

\title[]{Description of resonances within the rigged Hilbert space}

\classification{03.65.-w, 02.30.Hq}
\keywords{Resonances; Hilbert space; rigged Hilbert space; Gamow states}

\author{Rafael de la Madrid}{
  address={Department of Physics, University of 
           California at San Diego, La Jolla, CA 92093},
  altaddress={E-mail: \texttt{rafa@physics.ucsd.edu}}
}

\begin{abstract}
The spectrum of a quantum system has in general bound, scattering and 
resonant parts. The Hilbert space includes only the bound and scattering 
spectra, and discards the resonances. One must therefore enlarge the
Hilbert space to a rigged Hilbert space, within which the physical bound, 
scattering and resonance spectra are included on the same footing. In these 
lectures, I will explain how this is done.
\end{abstract}

\date{}

\maketitle

\section{Introduction: Lecture~1}
\label{sec:intro}

In Quantum Mechanics, observable quantities are represented by linear 
operators. The 
eigenvalues of an operator represent the possible values of the 
measurement of the corresponding observable. These eigenvalues, which 
mathematically correspond to the spectrum of the operator, can be discrete 
(as the energies of a particle in a box), continuous (as the energies 
of a free, unconstrained particle), resonant (as in $\alpha$ decay), or
a combination thereof.

The Hilbert space includes only the bound and scattering spectra, because the 
Hilbert space spectrum of an observable is real, thereby discarding the 
resonance spectrum as unphysical. However, radioactive nuclei and unstable
elementary particles are physical objects that ought to have a place in the 
quantum mechanical formalism. This is why we need to extend the Hilbert 
space to a rigged Hilbert space, within which the resonance spectrum has 
a place. 

The purpose of this series of lectures is to explain how one should use 
the rigged Hilbert space in quantum mechanics and, in particular, how
to incorporate the resonance spectrum into the quantum mechanical formalism 
by using the rigged Hilbert space.

When the spectrum of an observable $A$ is discrete and $A$ is 
bounded, then $A$ is defined on the whole of the Hilbert space $\cal H$ and
the eigenvectors of $A$ belong to $\cal H$. In this case, $A$ can be 
essentially seen as a matrix. This means that, as 
far as discrete spectrum is concerned, there is no need to extend 
$\cal H$. However, quantum mechanical
observables are in general unbounded and their spectrum has 
in general a continuous part. In order to deal with continuous 
spectrum, we use Dirac's bra-ket formalism. This formalism does not fit
within the Hilbert space alone, but within the rigged Hilbert space.

Loosely speaking, a rigged Hilbert space (also called a Gelfand triplet) is
a triad of spaces
\begin{equation}
       {\Phi} \subset {\cal H} \subset {\Phi}^{\times}
      \label{RHStIntro}
\end{equation}
such that $\cal H$ is a Hilbert space, $\Phi$ is a dense
subspace of $\cal H$, and $\Phi ^{\times}$ is the space of
antilinear functionals over $\Phi$. Mathematically,
$\Phi$ is the space of test functions, and $\Phi ^{\times}$
is the space of distributions. The space $\Phi ^{\times}$ is called
the antidual space of $\Phi$. Associated with the 
rigged Hilbert space~(\ref{RHStIntro}), there is always another rigged
Hilbert space,
\begin{equation}
       {\Phi} \subset {\cal H} \subset {\Phi}^{\prime} \, ,
      \label{RHSpIntro}
\end{equation}
where ${\Phi}^{\prime}$ is called the dual space of ${\Phi}$
and contains the linear functionals over $\Phi$. 

The basic reason why we need the spaces ${\Phi}^{\prime}$ and
${\Phi}^{\times}$ is that the bras and kets associated with the 
elements in the continuous spectrum of an observable belong, respectively, to 
${\Phi}^{\prime}$ and ${\Phi}^{\times}$ rather than to
${\cal H}$. The basic reason reason why we need the space $\Phi$ is 
that unbounded operators are not defined on the whole of ${\cal H}$ but only
on dense subdomains of ${\cal H}$ that are not invariant under the
action of the observables. Such non-invariance makes expectation values,
uncertainties and commutation relations not well defined on the whole
of $\cal H$. The space $\Phi$ is the largest subspace of the Hilbert 
space on which such expectation values, uncertainties and commutation 
relations are well defined.

Besides accommodating resonances and Dirac's bra-ket formalism, the
rigged Hilbert space seems to capture the physical principles of quantum
mechanics better than the Hilbert space. For example, assuming that the 
Hilbert space provides the whole mathematical framework for quantum mechanics 
leads to the conclusion that Heisenberg's uncertainty relations
are not physical, since they cannot be defined on the whole of the Hilbert 
space~\cite{PERES}. Using the rigged Hilbert space, one overcomes this 
difficulty after realizing that the commutation relations are well defined 
on $\Phi$.

The completeness relation is a good place to appreciate the added value of
the rigged Hilbert space. Consider, for example, the Hamiltonian $H$ of a 
system. In the Hilbert space, one writes the completeness relation as
\begin{equation}
      1=\int_{{\rm Sp}(H)}  d{\sf E}_E \, , 
\end{equation}
where ${\sf E}_E$ are the spectral projections of $H$ and ${\rm Sp}(H)$ is 
its spectrum. However, within the rigged Hilbert space one can 
write\footnote{In the Hilbert space, one can actually write something close
to, although not the same as~(\ref{crHS}), by means of direct integral
decompositions.}
\begin{equation}
      1= \sum_n |E_n\rangle \langle E_n| +
         \int_0^{\infty}dE \, |E\rangle \langle E| \, , 
         \label{crHS}
\end{equation}
where $|E_n\rangle$ and $|E\rangle$ are the bound and scattering states of 
$H$, respectively. In addition to~(\ref{crHS}), the rigged Hilbert space 
gives you an additional completeness relation in which the resonance states 
participate:
\begin{equation}
      1= \sum_n |E_n\rangle \langle E_n| +\sum_n |z_n\rangle \langle z_n|
         + \int_{-\infty}^0dE \, |E\rangle \langle E| \, ,
       \label{crRHS}
\end{equation}
where $|z_n\rangle$ are the Gamow (resonance) states of $H$ and the last
integral, called the background, is performed in the complex plane right 
below the negative real axis of the second sheet.\footnote{As we will
see in Lecture~5, the expansion~(\ref{crRHS}) must be either regulated or 
understood in a time-dependent way.} Thus, the
completeness relation~\eqref{crRHS} substitutes 
the scattering states contribution by the resonance contribution plus a 
background, thereby putting the resonance spectrum on the same footing as the
bound and scattering spectra. 

It is important to note that the integrals in (\ref{crHS}) and (\ref{crRHS}) 
are different, and that the resonance contribution does not appear
in~(\ref{crHS}), because resonances are not asymptotic states. Also important 
is to note that the resonance states, and therefore expansion~(\ref{crRHS}), 
need a different rigged Hilbert space from that needed by the scattering 
states and expansion~(\ref{crHS}).

There are dangers in using the rigged Hilbert space indiscriminately, 
though. For instance, A.~Bohm and collaborators have been using a
rigged Hilbert space (of Hardy class) to construct a quantum theory of 
resonances, see review~\cite{IJTP03} and references therein. However, such 
theory is inconsistent with quantum mechanics and must be 
discarded~\cite{HARDY}.

The structure of these lectures is as follows. After an introductory Lecture~1,
I will explain in Lecture~2 how to construct the rigged Hilbert space of the
one-dimensional rectangular barrier potential. This will show that the rigged 
Hilbert space is already needed to provide the mathematical support of the 
most basic quantum systems. In Lecture~3, I will construct the rigged Hilbert 
space of the Lippmann-Schwinger equation, which equation governs quantum
scattering. In Lecture~4, I will construct the rigged Hilbert
space of the analytic continuation of the Lippmann-Schwinger equation. The
rigged Hilbert space of Lecture~4 will be needed in the final Lecture~5 to 
provide the mathematical support for the resonance states. The PDF file
of each talk will be posted at \url{http://www.ucsd.edu/~rafa}.

Since space prevents a full account, at each lecture I will refer 
the reader to the appropriate papers where further details can be found. The 
essentials of functional analysis needed to understand those papers can be 
found in~\cite[Chapter 2]{DIS}.

\section{Construction of a simple rigged Hilbert space: Lecture~2}
\label{sec:motivRHS}

If resonances are not included, the way to construct the rigged Hilbert space
of a quantum system is as follows:
\begin{enumerate}
   \item We identify the observables of the system. Their expressions are 
usually given by linear, differential operators.
   \item We identify the Hilbert space, whose scalar product is used to
calculate probability amplitudes.
   \item We identify the domains, spectra and eigenfunctions of the 
observables. If the 
observables have a discrete spectrum, we need not go beyond the Hilbert 
space. However, if at least one of the observables has continuous spectrum, we
need to enlarge the Hilbert space to the rigged Hilbert space. (If position
and momentum are among the observables, we will
always need the rigged Hilbert space.)
   \item We construct the space $\Phi$ in which physical quantities such as
expectation values, uncertainties and commutation relations are well 
defined. When resonances are not included, the space $\Phi$ is usually 
given by the maximal invariant subspace of the algebra of observables.
   \item We construct the dual $\Phi '$ and antidual $\Phi ^{\times}$ 
spaces. We construct the bras and kets of the observables and 
check that they respectively belong to $\Phi '$ and $\Phi ^{\times}$.
   \item The completeness relations and all the features of Dirac's bra-ket 
formalism now follow.
\end{enumerate}

Let's see how the above steps are carried out in the case of a spinless 
particle moving in one dimension and impinging on a rectangular barrier. The 
observables relevant to this system are the position $Q$, the momentum $P$, and
the Hamiltonian $H$: 
\begin{equation}
      Qf(x)=xf(x) \, ,
        \label{fdopx}
\end{equation}
\begin{equation}
      Pf(x)=-i \hbar \frac{d}{d x}f(x) \, , 
        \label{fdopp}
\end{equation}
\begin{equation} 
      Hf(x)= \left( -\frac{\hbar ^2}{2m}\frac{d ^2}{d x^2}+V(x) \right) 
                  f(x) 
          \, , 
         \label{fdoph}
\end{equation}
where
\begin{equation}
           V(x)=\left\{ \begin{array}{ll}
                                0   &-\infty <x<a  \\
                                V_0 &a<x<b  \\
                                0   &b<x<\infty 
                  \end{array} 
                 \right. 
	\label{sbpotential1D}
\end{equation}
is the 1D rectangular barrier potential. These observables satisfy 
the following commutation relations:
\begin{equation}
      \left[ Q,P \right] =i \hbar I \, , \label{cr1} 
\end{equation}
\begin{equation}
      \left[ H,Q \right] =- \frac{i \hbar}{m} P \, ,   
\end{equation}
\begin{equation}
      \left[ H,P \right] = i \hbar \frac{\partial V}{\partial x} \, .  
        \label{cr3}
\end{equation}

Since our particle can move in the full real line, the Hilbert space
on which the differential operators~(\ref{fdopx})-(\ref{fdoph}) should act is 
\begin{equation}
     L^2 =\{ f(x) \, | \ \int_{-\infty}^{\infty}d x \, 
                  |f(x)|^2 < \infty  \}   \, .
     \label{l2space}
\end{equation}
The corresponding scalar product is
\begin{equation}
      (f,g)=\int_{-\infty}^{\infty}d x \, \overline{f(x)}g(x) \, , \qquad
      f,g \in L^2 \, .
          \label{scapro}
\end{equation}

The differential operators~(\ref{fdopx})-(\ref{fdoph})
induce three linear operators on the Hilbert space $L^2$. These operators 
cannot be defined on the whole of $L^2$, but only on the following 
subdomains of $L^2$:
\begin{equation}
     {\cal D}(Q)=\left\{ f\in L^2 \, | \ 
                   xf \in  L^2 \right\} \, ,  
\end{equation}
\begin{equation}
     {\cal D}(P)=\left\{ f\in L^2 \, | \ 
                f \in  AC, \
                Pf \in  L^2 \right\} \, , 
\end{equation}
\begin{equation}
     {\cal D}(H)=\left\{ f\in L^2 \, | \ 
                f \in  AC^2, \
                Hf \in  L^2 \right\} \, , 
         \label{domainH} 
\end{equation}
where, essentially, $AC$ is the space of functions whose derivative exists, 
and $AC^2$ is the space of functions whose second derivative exists. On these 
domains, the operators $Q$, $P$ and $H$ are self-adjoint, and their spectra are
\begin{equation}
      {\rm Sp}(Q)={\rm Sp}(P)=(-\infty, \infty) \, , \quad
     {\rm Sp}(H)=[0, \infty) \, , 
\end{equation}
which spectra coincide with those we would expect on physical grounds. 

To obtain the eigenfunctions corresponding to each eigenvalue, we have to 
solve the eigenvalue equation for each observable: 
\begin{equation}
      x \langle x|x'\rangle = x' \langle x|x'\rangle \, , 
         \label{eeq}
\end{equation}
\begin{equation} 
      -i \hbar \frac{d}{d x} \langle x|p\rangle = 
              p \langle x|p\rangle  \, ,
       \label{eep}
\end{equation}
\begin{equation} 
     \left( -\frac{\hbar ^2}{2m}\frac{d ^2}{d x^2}    
        +V(x) \right) \langle x|E\rangle = E \langle x|E\rangle \, . 
         \label{eeh}
\end{equation}
The eigenfunctions of $Q$ are delta functions,
\begin{equation}
      \langle x| x'\rangle = \delta (x-x') \, ,
       \label{deltax}
\end{equation}
those of $P$ are plane waves,
\begin{equation}
      \langle x| p\rangle = \frac{e ^{i px/\hbar}}{\sqrt{2\pi \hbar}} 
                   \, ,
       \label{expp}
\end{equation}
and those of $H$ are given by
\begin{equation}
           \langle x|E^+\rangle _{\rm r}   = 
          \left( \frac{m}{2\pi k \hbar ^2} \right)^{1/2} \times   
                \left\{ \begin{array}{lc}
             T (k)e^{-i kx}  &-\infty <x<a  \\
             A_{\rm r}(k)^{i \kappa x}+
                       B_{\rm r}(k) e^{-i \kappa x} &a<x<b \\
        R_{\rm r}(k)e^{i kx} + e ^{-i kx}   &b<x<\infty \, ,
                  \end{array} 
                 \right. 
     \label{chir+}
\end{equation}
\begin{equation}
         \langle x|E^+\rangle _{\rm l}=          
        \left( \frac{m}{2\pi k \hbar ^2} \right)^{1/2}
               \times  \left\{ \begin{array}{lc}
             e ^{i kx}+R_{\rm l}(k)e ^{-i kx}  &-\infty <x<a  \\
     A_{\rm l}(k)e ^{i \kappa x}+B_{\rm l}(k)e ^{-i \kappa x}
                                   &a<x<b \\
               T (k)e ^{i kx}         &b<x<\infty \, ,
                  \end{array} 
                 \right.
          \label{chil+} 
\end{equation}
where
\begin{equation}
      k=\sqrt{\frac{2m}{\hbar ^2}E} \, , \quad
      \kappa =\sqrt{\frac{2m}{\hbar ^2}(E-V_0)} \, ,
          \label{qwavnuem}
\end{equation}
and where the coefficients that appear in Eqs.~(\ref{chir+})-(\ref{chil+}) can 
be easily found by the standard matching conditions at the discontinuities of 
the potential. Physically, $\langle x|E^+\rangle _{\rm r}$
($\langle x|E^+\rangle _{\rm l}$) represents a particle of energy $E$ 
impinging on the barrier from the right (left).

The eigenfunctions (\ref{deltax})-(\ref{chil+}) are not square integrable, 
that is, they do not belong to $L^2$. Mathematically speaking, this is the 
reason why they are
to be dealt with as distributions (note that all of them except for the delta
function are also proper functions). 

We now start the construction of the rigged Hilbert space by constructing 
$\Phi$. The space $\Phi$ is given by
\begin{equation}
       \Phi =\bigcap _{\Sb n,m=0 \\ A,B=Q,P,H \endSb}^{\infty} 
            {\cal D}(A^nB^m) \, .
      \label{maximalinvas}
\end{equation}
In view of expressions (\ref{fdopx})-(\ref{fdoph}), $\Phi$ is simply
\begin{eqnarray}
      {\Phi} =\{ \varphi \in L^2 \, | \
    \varphi \in C^{\infty}(\mathbb R), \ \varphi ^{(n)}(a)=\varphi ^{(n)}(b)=0 
    \, , \ n=0,1,\ldots \, ,  \nonumber \\
    \hskip3.6cm
     P^nQ^mH^l\varphi (x) \in L^2
    \, , \   n,m,l=0,1, \ldots  \} \, ,
     \label{ddomain}
\end{eqnarray}
where $C^{\infty}(\mathbb R)$ is the collection of infinitely differentiable
functions, and $\varphi ^{(n)}$ denotes the $n$th derivative of 
$\varphi$. From the last condition in Eq.~(\ref{ddomain}), we deduce that the 
elements of $\Phi$ satisfy the following estimates:
\begin{equation}
      \| \varphi \| _{n,m,l} \equiv  
 \sqrt{\int_{-\infty}^{\infty}d x 
          \, \left| P^nQ^mH^l\varphi (x)\right| ^2 \,}
   < \infty  \, , \quad n,m,l=0,1,\ldots \, .
      \label{nmnorms}
\end{equation}
These estimates mean that the action of any combination of any power of the
observables remains square integrable. For this to happen, the functions
$\varphi (x)$ must be infinitely differentiable and must fall off at infinity
faster than any polynomial. Hence, $\Phi$ is a Schwartz-like space.

Because $\Phi$ is invariant under the action of the observables,
\begin{equation}
      A \, \Phi \subset \Phi \, , \qquad A=P,Q,H,
\end{equation}
the expectation values
\begin{equation}
      (\varphi , A^n\varphi ) \, , \quad \varphi \in \Phi \, ,
      \ A=P, Q, H , \ n=0,1,\ldots 
\end{equation}
are finite, and the commutation relations~(\ref{cr1})-(\ref{cr3}) are 
well defined. In particular, Heisenberg's uncertainty principle makes 
sense on $\Phi$.

The spaces $\Phi '$ and $\Phi ^{\times}$ are simply the collection of 
linear and antilinear functionals over $\Phi$, respectively. By combining 
the spaces $\Phi$, $\cal H$, $\Phi ^{\times}$ and $\Phi '$, we obtain the 
rigged Hilbert spaces of our system,
\begin{equation}
       \Phi \subset {\cal H}\subset \Phi ^{\times}
        \, ,
       \label{RHSCONT}
\end{equation}
\begin{equation}
       \Phi \subset {\cal H}\subset \Phi '
        \, .
       \label{RHSprime}
\end{equation}
The space $\Phi ^{\times}$ accommodates the eigenkets $|p \rangle$, 
$|x\rangle$ and $|E^+\rangle _{\rm l,r}$ of $P$, $Q$ and $H$, whereas
$\Phi '$ accommodates the eigenbras $\langle p|$, $\langle x|$ and 
$_{\rm l,r}\langle ^+E|$.

Mathematically, the bras and kets are distributions defined as follows. Given 
a function $f(x)$ and a space 
of test functions $\Phi$, the antilinear functional $F$ that 
corresponds to the function $f(x)$ is an integral operator whose kernel is 
precisely $f(x)$:
\begin{equation}
      F(\varphi) \equiv \int d x \, \overline{\varphi (x)} f(x) \, ,
         \label{afFtff1}
\end{equation}
and the linear functional $\tilde{F}$ generated by the function $f(x)$ is 
an integral operator whose kernel is the complex conjugate of $f(x)$:
\begin{equation}
      \tilde{F}(\varphi)\equiv \int d x \, \varphi (x) \overline{f(x)} \, .
          \label{afGtff1}
\end{equation}
In Dirac's notation, these two equations become
\begin{equation}
      \langle \varphi|F\rangle =\int d x \, \langle \varphi|x\rangle 
                                     \langle x|f\rangle \, , 
       \label{afFtff2}
\end{equation}
\begin{equation}
      \langle F|\varphi \rangle =\int d x \, \langle f|x\rangle
                                            \langle x|\varphi \rangle \, . 
\end{equation}
Note that these definitions are very similar, except that the complex 
conjugation affects either $f(x)$ or $\varphi (x)$, which makes the 
corresponding functional either linear or antilinear.

Definitions~(\ref{afFtff1}) and (\ref{afGtff1}) provide the link between the 
quantum mechanical formalism and the theory of distributions. In practical 
applications, what one obtains from the quantum mechanical formalism is the 
distribution $f(x)$ (in this lecture, the plane waves
$\frac{1}{\sqrt{2\pi \hbar}} e ^{i px/\hbar}$, the delta function
$\delta (x-x')$ and the eigenfunctions 
$\langle x|E^+\rangle _{\rm l,r}$). Once $f(x)$ is given, one
can use definitions~(\ref{afFtff1}) and (\ref{afGtff1}) to generate the 
functionals $|F\rangle$ and $\langle F|$. Then,
the theory of distributions can be used to obtain the properties of the 
functionals $|F\rangle$ and $\langle F|$, which in turn yield the properties 
of the distribution $f(x)$. 

By using prescription~(\ref{afFtff1}), we can define for 
each eigenvalue $p$ the eigenket $|p\rangle$ associated with the 
eigenfunction (\ref{expp}):
\begin{equation}
       \langle \varphi |p\rangle  \equiv 
       \int_{-\infty}^{\infty}d x \, \overline{\varphi (x)}
          \frac{1}{\sqrt{2\pi \hbar}} e ^{i px/\hbar}  \, ,
     \label{definitionketp}
\end{equation}
which, using Dirac's notation for the integrand, becomes 
\begin{equation}
       \langle \varphi |p\rangle  \equiv 
       \int_{-\infty}^{\infty}d x \, \langle \varphi |x\rangle 
       \langle x|p\rangle \, .
     \label{definitionketpDirac}
\end{equation}
Similarly, for each $x$, we can define the ket $|x\rangle$ as
\begin{equation}
     \langle \varphi |x\rangle  \equiv 
       \int_{-\infty}^{\infty}d x' \, \overline{\varphi (x')}
          \delta (x-x') \, ,
     \label{definitionketx}
\end{equation}
which, using Dirac's notation for the integrand, becomes
\begin{equation}
     \langle \varphi |x\rangle  \equiv 
       \int_{-\infty}^{\infty}d x' \, \langle \varphi | x'\rangle
          \langle x'|x\rangle \, .
     \label{definitionketxDirac}
\end{equation}
The definition of the kets $|E^+\rangle _{\rm l,r}$ that correspond to the 
Hamiltonian's eigenfunctions~(\ref{chir+})-(\ref{chil+})
follows the same prescription:
\begin{equation}
     \langle \varphi |E^+\rangle _{\rm l,r}  \equiv 
       \int_{-\infty}^{\infty}d x \, \overline{\varphi (x)}
        \langle x|E^+\rangle _{\rm l,r}  \, ,
    \label{definitionketE}
\end{equation}
that is, 
\begin{equation}
     \langle \varphi |E^+\rangle _{\rm l,r}  \equiv 
       \int_{-\infty}^{\infty}d x \, \langle \varphi |x\rangle
          \langle x|E^+\rangle _{\rm l,r} \, .
    \label{definitionketEDirac}
\end{equation}
One can now show that the definition of the kets $|p\rangle$, 
$|x\rangle$ and $|E^+\rangle _{\rm l,r}$ makes sense, and that
these kets indeed belong to the space of distributions $\Phi ^{\times}$.

By using prescription~(\ref{afGtff1}), we can also define for each eigenvalue
$p$ the eigenbra $\langle p|$ associated with the eigenfunction~(\ref{expp}):
\begin{equation}
    \langle p| \varphi \rangle  \equiv 
       \int_{-\infty}^{\infty}d x \, \varphi (x)
          \frac{1}{\sqrt{2\pi \hbar}} e ^{-i px/\hbar}
         \equiv 
       \int_{-\infty}^{\infty}d x \, \langle p|x \rangle 
        \langle x| \varphi \rangle  \, .
     \label{definitionbrap}
\end{equation}
Comparison with Eq.~(\ref{definitionketp}) shows that the action of
$\langle p|$ is the complex conjugate of the action of $|p \rangle$,
\begin{equation}
      \langle p| \varphi \rangle = \overline{\langle \varphi |p \rangle} \, ,
\end{equation} 
and that
\begin{equation}
      \langle p| x \rangle = \overline{\langle x |p \rangle} =
      \frac{1}{\sqrt{2\pi \hbar}} e ^{-i px/\hbar}  \, .
    \label{pxiscmplxp}
\end{equation} 
The bra $\langle x|$ is defined as
\begin{equation}
     \langle x| \varphi \rangle  \equiv 
       \int_{-\infty}^{\infty}d x' \, \varphi (x')
          \delta (x-x') \equiv 
       \int_{-\infty}^{\infty}d x' \, \langle x | x'\rangle
          \langle x'| \varphi \rangle \, .
     \label{definitionbrax}
\end{equation}
Comparison with Eq.~(\ref{definitionketx}) shows that the action of
$\langle x|$ is complex conjugated to the action of $|x \rangle$,
\begin{equation}
      \langle x| \varphi \rangle = \overline{\langle \varphi |x \rangle} \, ,
\end{equation} 
and that
\begin{equation}
      \langle x| x' \rangle = \langle x' |x \rangle = \delta (x-x')  \, .
\end{equation} 
Analogously, the eigenbras of the Hamiltonian are defined as
\begin{equation}
     _{\rm l,r}\langle ^+E|\varphi\rangle \equiv
       \int_{-\infty}^{\infty}d x \ \varphi (x) \
           _{\rm l,r} \langle ^+E|x\rangle \equiv
       \int_{-\infty}^{\infty}d x \ 
       _{\rm l,r}\langle ^+E|x\rangle \langle x|\varphi \rangle   \, ,
    \label{definitionbraE}
\end{equation}
where
\begin{equation}
     _{\rm l,r}\langle ^+E|x\rangle = 
   \overline{\langle x|E^+\rangle}_{\rm l,r} \, .
\end{equation}
Comparison of Eq.~(\ref{definitionbraE}) with 
Eq.~(\ref{definitionketE}) shows that 
the actions of the bras $_{\rm l,r}\langle ^+E|$ are the complex 
conjugates of the actions of the kets $|E ^+ \rangle _{\rm l,r}$:
\begin{equation}
      _{\rm l,r}\langle ^+E|\varphi\rangle =      
      \overline{\langle \varphi |E ^+ \rangle}_{\rm l,r} \, .
      \label{braketccE}
\end{equation} 
Now, by using the rigged Hilbert space, one can show that the definitions of 
$\langle p|$, $\langle x|$ and $_{\rm l,r} \langle ^+E|$ make sense
and that $\langle p|$, $\langle x|$ and $_{\rm l,r} \langle ^+E|$
belong to $\Phi '$.

It is important to keep in mind the difference between eigenfunctions and
kets. For instance, $\langle x|p\rangle$ is an eigenfunction
of a differential equation, Eq.~(\ref{eep}), whereas $|p\rangle$ is a 
functional, the relation between them being given by 
Eq.~(\ref{definitionketpDirac}). A similar relation holds between 
$\langle x'|x\rangle$ and $|x\rangle$, and between 
$\langle x|E^+\rangle _{\rm l,r}$ and $|E^+\rangle _{\rm l,r}$. It
is also important to keep in mind that ``scalar products'' like
$\langle x|p\rangle$, $\langle x'|x\rangle$ or 
$\langle x|E^+\rangle _{\rm l,r}$ do not represent an actual scalar 
product of two functionals; these ``scalar products'' are simply solutions
to differential equations.

The kets $|p\rangle$, $|x\rangle$ and $|E^+\rangle_{\rm l,r}$ are 
indeed eigenvectors of $P$, $Q$ and $H$, respectively:
\begin{equation}
       P|p\rangle=p|p\rangle \, , \quad p\in \mathbb R \, ,
        \label{kpeP}
\end{equation}
\begin{equation}
       Q|x\rangle=x|x\rangle \, , \quad x \in \mathbb R \, ,
        \label{kxeQ}
\end{equation}
\begin{equation}
       H|E^+\rangle_{\rm l,r} =E|E^+\rangle _{\rm l,r}
   \, , \quad E\in [0,\infty )  \, .
        \label{kEeH}
\end{equation}
Similarly, the bras $\langle p|$, 
$\langle x|$ and $_{\rm l,r} \langle ^+E|$ are left eigenvectors of $P$, $Q$ 
and $H$, respectively:
\begin{equation}
       \langle p|P=p\langle p| \, , \quad p\in \mathbb R \, ,
        \label{bpeP}
\end{equation}
\begin{equation}
      \langle x|Q=x\langle x| \, , \quad x \in \mathbb R \, ,
\end{equation}
\begin{equation}
       _{\rm l,r} \langle ^+E|H=
       E \hskip0.12cm  
         _{\rm l,r} \langle ^+E|    \, , \quad E\in [0,\infty )  \, .
        \label{bEeH}
\end{equation}
Note that these equations are to be understood in the distributional way,
that is, as ``sandwiches'' with elements of $\Phi$. For example, 
Eq.~(\ref{kpeP}) should be understood as
\begin{equation}
       \langle \varphi |P|p\rangle=
           p  \langle \varphi|p\rangle \, , \quad p\in \mathbb R  \, ,
               \  \varphi \in \Phi \, ,
        \label{kpePsan}
\end{equation}
and the same for~(\ref{kxeQ})-(\ref{bEeH}). Usually, the ``sandwiching''
is implicit and therefore omitted. 

Now that we have constructed the Dirac bras and kets, we can
see how other aspects of Dirac's bra-ket formalism hold within the
rigged Hilbert space. For example, the completeness relations
\begin{equation}
      \int_{-\infty}^{\infty}  d p \, |p\rangle \langle p| = I \, ,
      \label{resonidentP}
\end{equation}
\begin{equation}
      \int_{-\infty}^{\infty}  d x' \, |x'\rangle \langle x'| = I \, ,
        \label{resonidentQ}
\end{equation}
\begin{equation}
      \int_{0}^{\infty} d E \, |E^+\rangle _{\rm l}\, 
                       _{\rm l}\langle ^+E| +
      \int_{0}^{\infty}d E \, |E^+\rangle _{\rm r}\, 
                       _{\rm r}\langle ^+E| =  I \, ,
        \label{resonidentH}
\end{equation}
and the action of $P$, $Q$ and $H$,
\begin{equation}
      P = \int_{-\infty}^{\infty}d p \, p |p\rangle  \langle p| \, ,
      \label{presAP}
\end{equation} 
\begin{equation}
      Q = \int_{-\infty}^{\infty}d x \, x |x\rangle  \langle x| \, ,
      \label{presAQ}
\end{equation} 
\begin{equation}
      H = \int_{0}^{\infty}d E \, E |E^+\rangle_{\rm l}\, 
       _{\rm l}\langle ^+E| +
        \int_{0}^{\infty}d E \, E |E^+\rangle_{\rm r}\, 
       _{\rm r}\langle ^+E| \, ,
     \label{presAH}
\end{equation} 
all hold within the rigged Hilbert space as a ``sandwich'' with elements
of $\Phi$. Other expressions such as the delta normalization of 
eigenfunctions, e.g.,
\begin{equation}
      \frac{1}{2\pi \hbar} \int_{-\infty}^{\infty}  
         d x \, e^{i(p-p')x/\hbar} = \delta (p-p') \, ,
           \label{pppdeltappp3}
\end{equation}
or the ``matrix elements'' of the observables, e.g.,
\begin{equation}
        \langle x|Q|x'\rangle = x' \, \delta (x-x') \, ,
         \label{meQxx}
\end{equation}
\begin{equation}
        \langle x|P|x'\rangle = -i \hbar \frac{d}{d x} \ 
                       \delta (x-x') \, ,
        \label{mePxx}
\end{equation}
\begin{equation}
        \langle x|H|x'\rangle = 
    \left( -\frac{\hbar ^2}{2m}\frac{d ^2}{d x^2}+V(x) \right) 
            \delta (x-x')   \, .
     \label{meHxx}
\end{equation}
are interpreted the same way.

To conclude this section, I would like to refer the reader 
to~\cite{04JPA,05EJP} for a detailed account of the above results.

\section{The rigged Hilbert space of the Lippmann-Schwinger equation: 
Lecture~3}
\label{sec:RHSLSEq}

The Lippmann-Schwinger equation is one of the cornerstones of scattering
theory. It is written as
\begin{equation}
       |E ^{\pm}\rangle =|E\rangle +
       \frac{1}{E-H_0\pm i \epsilon}V|E^{\pm}\rangle \, ,
       \label{LSeq1}
\end{equation}
where $|E^{\pm}\rangle$ are the ``in'' and ``out'' Lippmann-Schwinger 
kets, $|E\rangle$ is an eigenket of the free Hamiltonian $H_0$,
\begin{equation}
      H_0|E\rangle =E|E\rangle \, ,
      \label{tisequa0}
\end{equation}
and $V$ is the potential. The Lippmann-Schwinger kets are, in particular,
eigenvectors of $H$:
\begin{equation}
      H|E^{\pm}\rangle =E|E ^{\pm}\rangle \, .
      \label{tisequa}
\end{equation}
To the kets $|E^{\pm}\rangle$, there correspond the bras $\langle ^{\pm}E|$, 
which satisfy
\begin{equation}
        \langle ^{\pm}E| = 
        \langle E| +\langle ^{\pm}E|V\frac{1}{E-H_0\mp i \epsilon} \, .
       \label{LSeqbraspm}
\end{equation}
The bras $\langle ^{\pm}E|$ are left eigenvectors of $H$,
\begin{equation}
        \langle ^{\pm}E|H = E \langle ^{\pm}E|  \, ,
       \label{LSeibenbraeq}
\end{equation}
and the bras $\langle E|$ are left eigenvectors of $H_0$,
\begin{equation}
        \langle E|H_0 = E \langle E|  \, .
       \label{LSeibenbraeq0}
\end{equation}

The Lippmann-Schwinger equation~(\ref{LSeq1}) for the ``in''
$|E^{+}\rangle$ and ``out'' $|E^{-}\rangle$ kets has the scattering ``in'' 
and ``out'' boundary conditions built into the $\pm i \epsilon$, since 
Eq.~(\ref{LSeq1}) is equivalent to the time-independent Schr\"odinger 
equation~(\ref{tisequa}) subject to those ``in'' ($+i \epsilon$) and 
``out'' ($-i \epsilon$) boundary conditions. In the position 
representation, the $\pm i \epsilon$ prescriptions yield the following 
asymptotic behaviors:
\begin{equation}
      \langle {\bf x}|E^+\rangle \, \mapupdown{}{r\to \infty} \
      e^{i kz}+f(k,\theta) \, \frac{e^{i kr}}{r} \, ,
       \label{asympbepk}
\end{equation}
\begin{equation}
      \langle {\bf x}|E^-\rangle \, \mapupdown{}{r \to \infty} \
      e ^{i kz} + \overline{f(k,\theta )} \, 
           \frac{e ^{-i kr}}{r} \, ,
      \label{asympbemk}
\end{equation}
where ${\bf x}\equiv (x,y,z) \equiv (r,\theta , \phi)$ are the position
coordinates, $k$ is the wave number of Eq.~(\ref{qwavnuem}) and 
$f(k, \theta )$ is the scattering amplitude. 

Likewise any bra and ket, the Lippmann-Schwinger bras and kets do not have
a place in the Hilbert space. In this lecture, we will construct the rigged 
Hilbert space to which they belong. We will use the example of the spherical 
shell potential.

\subsection{The radial Lippmann-Schwinger equation}

For the spherical shell potential, 
\begin{equation}
        V({\bf x})\equiv V(r)=\left\{ \begin{array}{ll}
                                0   &0<r<a  \\
                                V_0 &a<r<b  \\
                                0   &b<r<\infty \, ,
                  \end{array} 
                 \right. 
	\label{sbpotential}
\end{equation}
the Lippmann-Schwinger equation can be solved explicitly. Because 
the potential~(\ref{sbpotential}) is spherically symmetric, we will work 
in the radial position representation and restrict ourselves to angular 
momentum $l=0$.

In the radial representation, the Lippmann-Schwinger equation~(\ref{LSeq1}) 
becomes
\begin{equation}
       \langle r|E ^{\pm}\rangle =\langle r|E\rangle +
       \langle r|\frac{1}{E-H_0\pm i \epsilon}V|E^{\pm}\rangle \, .
       \label{raLSeq}
\end{equation}
The procedure to solve Eq.~(\ref{raLSeq}) is well known. Since 
Eq.~(\ref{raLSeq}) is an integral equation,
it is equivalent to a differential equation subject to the boundary
conditions that are built into it. In our case,
for $l=0$, Eq.~(\ref{raLSeq}) is equivalent to the Schr\"odinger 
differential equation,
\begin{equation}
      \left( -\frac{\hbar ^2}{2m} \frac{d ^2}{d r^2}+V(r)\right) 
      \langle r|E^{\pm}\rangle =E\langle r|E^{\pm}\rangle \, , 
      \label{schroeque}
\end{equation}
subject to the following boundary conditions:
\begin{eqnarray}
      &&\langle r|E^{\pm}\rangle =0 \, ,  \label{bcoLS1} \\
      &&\langle r|E^{\pm}\rangle {\rm \ is \ continuous \ at \ } r=a, b \, , \\
      &&\frac{d}{d r}\langle r|E^{\pm}\rangle 
             {\rm \ is \ continuous \ at \ } r=a, b 
         \, , \\
      && \langle r|E^+\rangle \sim 
          e^{-i kr}- S(E) \, e ^{i kr} 
                     \quad {\rm as \ } r\to \infty \, ,
        \label{bcoLS5}\\
      && \langle r|E^-\rangle \sim
         e^{i kr} - \overline{S(E)} \,  e^{-i kr} 
             \quad {\rm as \ } r\to \infty \, ,
        \label{bcoLS6}
\end{eqnarray}
where $S(E)$ is the $S$~matrix in the energy representation. The boundary 
conditions~(\ref{bcoLS5}) and~(\ref{bcoLS6}) originate from the 
$\pm i \epsilon$ conditions of Eq.~(\ref{raLSeq}). The asymptotic 
behaviors~(\ref{bcoLS5}) and~(\ref{bcoLS6}) are the
$l=0$, radial counterparts of the asymptotic behaviors~(\ref{asympbepk}) and
(\ref{asympbemk}).

If we insert (\ref{sbpotential}) into (\ref{schroeque}), and solve 
(\ref{schroeque}) subject to~(\ref{bcoLS1})-(\ref{bcoLS6}), we obtain
\begin{equation}
      \langle r|E^{\pm}\rangle \equiv 
       \chi ^{\pm}(r;E)= N(E) \,  
     \frac{\chi (r;E)}{{\cal J}_{\pm}(E)}\, , \qquad E\in [0,\infty ) \, ,
      \label{pmeigndu}
\end{equation}
where $N(E)$ is a delta-normalization factor,
\begin{equation}
      N(E)=\sqrt{\frac{1}{\pi}
         \frac{2m/\hbar ^2}{\sqrt{2m/\hbar ^2 \, E \,}\,}\,} \, ,
        \label{Nfactor}
\end{equation}
$\chi (r;E)$ is the so-called regular solution of Eq.~(\ref{schroeque}),
\begin{equation}
     \hskip-1cm  \chi (r;E) = \left\{
             \begin{array}{ll}
                  \sin ( \sqrt{ \frac{2m}{\hbar^2}E\,}\,r)  & 0<r<a \\
               {\cal J}_1(E)e^{i \sqrt{\frac{2m}{\hbar^2}(E-V_0)\,}\,r}
              +{\cal J}_2(E)e^{-i\sqrt{\frac{2m}{\hbar^2}(E-V_0)\,}\,r}
                                                         & a<r<b \\ 
                  {\cal J}_3(E) e^{i \sqrt{\frac{2m}{\hbar^2}E\,}\,r}
                   +{\cal J}_4(E) e^{-i\sqrt{\frac{2m}{\hbar^2}E\,}\,r}
                                                         & b<r<\infty \, ,
               \end{array}
              \right.
        \label{LSchi}
\end{equation}
and ${\cal J}_{\pm}(E)$ are the Jost functions,
\begin{equation}
        {\cal J}_+(E)=-2i {\cal J}_4(E) \, ,
\end{equation}
\begin{equation}
        {\cal J}_-(E)=2i {\cal J}_3(E) \, .
\end{equation}
The explicit expressions for ${\cal J}_1$-${\cal J}_4$ can be obtained by 
matching the values of $\chi (r;E)$ and of its derivative at the 
discontinuities of the potential. In terms of the Jost functions, the 
$S$ matrix is given by
\begin{equation}
    S(E)=\frac{{\cal J}_-(E)}{{\cal J}_+(E)} \, , \qquad E\in [0,\infty ) \, .
\end{equation}

Similarly to the ``right'' ones, we can obtain the ``left'' 
Lippmann-Schwinger eigenfunctions by writing Eq.~(\ref{LSeqbraspm}) in the 
radial position representation, 
\begin{equation}
       \langle ^{\pm}E| r \rangle =\langle E|r \rangle +
       \langle ^{\pm}E| V\frac{1}{E-H_0\mp i \epsilon}|r \rangle \, .
       \label{raLSeqbra}
\end{equation}
Solving Eq.~(\ref{raLSeqbra}) is analogous to solving Eq.~(\ref{raLSeq}). The 
solutions to~(\ref{raLSeqbra}) are the complex conjugates of the solutions 
to~(\ref{raLSeq}): 
\begin{equation}
      \langle ^{\pm}E|r\rangle = \overline{\chi ^{\pm}(r;E)} =        
          \chi ^{\mp}(r;E)  \, .
      \label{pmeigndubra}
\end{equation}

\subsection{The rigged Hilbert space of the Lippmann-Schwinger bras and kets}

By using~(\ref{afFtff1}), we associate ``in'' and ``out'' kets 
$|E^{\pm} \rangle$ with the eigenfunctions $\chi ^{\pm}(r;E)$ for each 
$E\in [0,\infty )$:
\begin{equation}
       \langle \varphi ^{\pm}|E^{\pm}\rangle \equiv
       \int_0^{\infty}d r \, \overline{\varphi ^{\pm}(r)} \chi ^{\pm}(r;E)  
       \equiv
       \int_0^{\infty}d r \, \langle \varphi ^{\pm}|r\rangle 
      \langle r|E^{\pm}\rangle  \, ,
        \label{LSdefinitionket+E}
\end{equation}
where $\varphi ^{\pm}$ belong to the space $\Phi$ that will be constructed 
below. (Note that the $\Phi$ of
this lecture is different from the $\Phi$ of Lecture~2.) Similarly,
by using~(\ref{afGtff1}), we define the bras $\langle ^{\pm}E|$ as
\begin{equation}
     \langle ^{\pm}E| \varphi ^{\pm}\rangle \equiv
       \int_0^{\infty} d r \, \overline{\chi^{\pm}(r;E)} \varphi ^{\pm}(r)
     \equiv
       \int_0^{\infty} d r \, 
              \langle ^{\pm}E|r\rangle \langle r|\varphi ^{\pm}\rangle \, .
      \label{defibraEplus}
\end{equation}
Note that even though $\chi ^{\pm}(r;E) \equiv \langle r|E^{\pm}\rangle$ 
are also meaningful for complex energies, the energy in 
Eqs.~(\ref{LSdefinitionket+E}) and (\ref{defibraEplus}) runs 
only over ${\rm Sp}(H)=[0,\infty )$, because in this lecture we restrict 
ourselves to bras and kets associated with energies that belong to the 
scattering spectrum of the Hamiltonian. 

From definitions~(\ref{LSdefinitionket+E}) and (\ref{defibraEplus}), it
follows that the action of $\langle ^{\pm}E|$ is complex conjugated 
to the action of $|E^{\pm} \rangle$:
\begin{equation}
      \langle ^{\pm}E| \varphi ^{\pm} \rangle =
       \overline{\langle \varphi ^{\pm}|E^{\pm} \rangle}  \, .
        \label{baccka}
\end{equation}

We now need to find the subspace $\Phi$ on which the above
definitions make sense. Besides making 
(\ref{LSdefinitionket+E})-(\ref{defibraEplus}) well defined, the space 
${\Phi}$ must also be invariant under the action of the observables of the 
system. Since in this lecture the only 
observable we are concerned with is the Hamiltonian, we will simply require 
invariance under $H$. Thus, the space $\Phi$ must satisfy the 
following conditions:
\begin{eqnarray}
  \hskip-1.7cm  &&\bullet \ \, {\rm The \ space \ } {\Phi} \ {\rm is 
      \ invariant \
      under \ the \ action \ of } \  H. 
                   \label{condition1} \\ [1ex]
    \hskip-1.7cm  &&\bullet \ \, {\rm The \ elements \ of \  } {\Phi} 
    \ {\rm are \ such \ that \ the \ integrals \ in \ 
    Eqs.~(\ref{LSdefinitionket+E}) \frac{\ \, }{\ \, }(\ref{defibraEplus}) 
             \ make \ sense.}
              \label{condition2} 
\end{eqnarray}

In order to meet requirement~(\ref{condition1}), the wave functions 
$\varphi ^{\pm}(r)$ must at least be in the maximal invariant subspace of $H$:
\begin{equation}
      {\cal D} = \bigcap_{n=0}^{\infty} {\cal D}(H^n) \, .
         \label{misoH}
\end{equation}
In order to 
meet requirement~(\ref{condition2}), the wave functions $\varphi ^{\pm}(r)$ 
must behave well enough so the integrals in 
Eqs.~(\ref{LSdefinitionket+E})-(\ref{defibraEplus}) are well defined. From the 
expression for $\chi ^{\pm}(r;E)$, 
Eq.~(\ref{pmeigndu}), one can see that the $\varphi ^{\pm}(r)$ 
have essentially to control purely imaginary exponentials. Therefore, the 
space $\Phi$ that meets the 
requirements~(\ref{condition1})-(\ref{condition2}) is
\begin{equation}
     \hskip-1.3cm  {\Phi}= 
      \left\{ \varphi ^{\pm}\in L^2([0,\infty ),d r) \, | \
        \varphi ^{\pm}\in {\cal D} \, , 
      \ \| \varphi ^{\pm} \|_{n,m}<\infty \, , \ n,m=0,1,2,\ldots \right\}  ,
         \label{spacePhi}
\end{equation}
where the $\| \ \|_{n,m}$ are given by
\begin{equation}
     \| \varphi ^{\pm} \|_{n,m} = \sqrt{\int_{0}^{\infty}d r \, 
    \left| (1+r)^n (1+H)^m \varphi ^{\pm}(r) 
      \right|^2 \, } \, , \quad n,m=0,1,2, \ldots \, .
      \label{normsLS}
\end{equation}
The space $\Phi$ is thus the collection of square integrable functions 
that belong to the maximal invariant subspace of $H$ and for which the 
estimates~(\ref{normsLS}) are finite. In particular, because 
$\varphi ^{\pm}(r)$ 
satisfy the estimates~(\ref{normsLS}), $\varphi ^{\pm}(r)$ fall off at
infinity faster than any polynomial of $r$: 
\begin{equation}
      \lim_{r\to \infty} (1+r)^n\varphi ^{\pm}(r) =0 \, , 
                \quad n=0,1,2,\ldots \, .
       \label{zeroinflimit}
\end{equation}
Thus, likewise in Lecture~2, we have obtained a Schwartz-like space. Obviously,
the space ${\Phi}$ can also be seen
as the maximal invariant subspace of the algebra generated by the Hamiltonian
and the operator multiplication by $r$. 

Note that we have used superscripts
$\pm$ to denote the elements of one and the same space $\Phi$. The reason is
that when we use $+$, it means that the elements of $\Phi$ are acted upon
by $|E^+\rangle$ or $\langle ^+E|$, and when we use $-$, they are acted upon
by $|E^-\rangle$ or $\langle ^-E|$.

We can now construct the spaces $\Phi '$ and $\Phi ^{\times}$, and see that
$|E^{\pm}\rangle$ belong to $\Phi ^{\times}$, whereas $\langle ^{\pm}E|$ 
belong to $\Phi '$. Thus, the Lippmann-Schwinger 
equations~(\ref{LSeq1}) and~(\ref{LSeqbraspm}), Eqs.~(\ref{tisequa}) and 
(\ref{LSeibenbraeq}), as well as
\begin{equation}
      e^{-iHt/\hbar}|E^{\pm}\rangle = e^{-iEt/\hbar}|E^{\pm}\rangle \, ,
\end{equation}
\begin{equation}
      \langle ^{\pm}E| e^{-iHt/\hbar} = e^{iEt/\hbar}\langle ^{\pm}E| \, ,
\end{equation}
hold in the distributional sense, i.e., as ``sandwiches'' with elements of 
$\Phi$. 

For a detailed account on the results of this lecture, the reader may
wish to refer to~\cite{LS1}.

\section{The rigged Hilbert space of the analytic continuation of the the 
Lippmann-Schwinger equation: Lecture~4}
\label{sec:RHSacLSe}

This lecture is devoted to construct and characterize
the analytic continuation of the Lippmann-Schwinger bras and
kets. As in Lecture~3, we restrict ourselves to the spherical shell 
potential~(\ref{sbpotential}) and zero angular momentum. 

The ultimate goal we want to achieve by analytically continuing the solutions 
of the Lippmann-Schwinger equation is to obtain, in Lecture~5, the resonance 
(decay) amplitude.

\subsection{The wave number representation}

The Lippmann-Schwinger eigenfunctions depend explicitly on 
the wave number $k$ of Eq.~(\ref{qwavnuem})
rather than on the energy $E$. It is therefore
convenient to rewrite their expressions in terms of $k$ before performing 
analytic continuations.

We start by writing the regular solution~(\ref{LSchi}) in terms of $k$:
\begin{equation}
      \chi (r;k)=  \chi (r;E) = \left\{ \begin{array}{lll}
                   \sin (kr) \quad &0<r<a  \\
                {\cal J}_1(k) e^{i \kappa r}+
                {\cal J}_2(k) e^{-i \kappa r}  \quad  &a<r<b \\
               {\cal J}_3(k)  e^{i kr}+{\cal J}_4(k) e^{-i kr}
                                  \quad  &b<r<\infty \, , 
               \end{array} 
                 \right.   
        \label{chiwna}
\end{equation}
where $\kappa$ is given by~(\ref{qwavnuem}). In terms of $k$, the 
Lippmann-Schwinger eigenfunctions read as
\begin{equation}
     \chi ^{\pm}(r;E)= \sqrt{\frac{1}{\pi} \frac{2m/\hbar ^2}{k}\,} \,  
     \frac{\chi (r;k)}{{\cal J}_{\pm}(k)} \, .
       \label{krepmeigndu}
\end{equation}
The eigenfunctions $\chi ^{\pm}(r;E)$ are $\delta$-normalized as functions
of $E$. The Lippmann-Schwinger eigenfunctions that are $\delta$-normalized 
as functions of $k$ are given by
\begin{equation}
      \chi ^{\pm}(r;k)\equiv \sqrt{\frac{\hbar ^2}{2m}2k\,}\, \chi ^{\pm}(r;E)=
                       \sqrt{\frac{2}{\pi}\,}\, 
                      \frac{\chi (r;k)}{{\cal J}_{\pm}(k)} \, .
      \label{defiphi+-}
\end{equation}
In bra-ket notation, we will write
\begin{equation}
      \langle r|k^{\pm}\rangle = \chi ^{\pm}(r;k)  \, , \quad k>0 \, , 
\end{equation}
\begin{equation}
      \langle ^{\pm}k|r \rangle = \overline{\chi ^{\pm}(r;k)} =     
            \chi ^{\mp}(r;k)      \, , \quad k>0 \, . 
     \label{leftLSe+-}
\end{equation}

Because of~(\ref{defiphi+-}), in terms of $k$ the Lippmann-Schwinger bras 
and kets read as
\begin{eqnarray}
      \langle ^{\pm}k| = \sqrt{\frac{\hbar ^2}{2m}\, 2k\,} \, \langle ^{\pm}E|
          \, , \quad k>0 \, ,  \label{brask} \\
      |k^{\pm}\rangle = \sqrt{\frac{\hbar ^2}{2m}\, 2k\,} \, |E^{\pm}\rangle 
          \, , \quad k>0 \, . 
          \label{ketsk}
\end{eqnarray}
The bras $\langle ^{\pm}k|$ and 
kets $|k^{\pm}\rangle$ are, respectively, left and right eigenvectors of $H$ 
with eigenvalue $\frac{\hbar ^2}{2m}k^2$:
\begin{eqnarray}
       \langle ^{\pm}k| H =\frac{\hbar ^2}{2m}k^2  \langle ^{\pm}k| \, , \\
       H |k^{\pm}\rangle =\frac{\hbar ^2}{2m}k^2  |k^{\pm}\rangle \, .
\end{eqnarray}

\subsection{The analytic continuation of the Lippmann-Schwinger eigenfunctions}

The analytic continuation of $\chi ^{\pm}(r;k)$ is done as follows. First, 
one specifies
the boundary values that the Lippmann-Schwinger eigenfunctions take on the 
positive $k$-axis. And second, one continues those boundary values into 
the whole $k$-plane. Since the boundary values of the Lippmann-Schwinger 
eigenfunctions on the positive $k$-axis are given by Eq.~(\ref{defiphi+-}),
and since $\chi ^{\pm}(r;k)$ are expressed in terms of well-known analytic 
functions, the continuation of $\chi ^{\pm}(r;k)$ from the positive $k$-axis 
into the whole wave-number plane is well defined.

A word on notation. Whenever they become complex, we will
denote the energy $E$ and the wave number $k$ by respectively $z$ and 
$q$. Accordingly, the
continuations of $\chi ^{\pm}(r;E)$ and $\chi ^{\pm}(r;k)$ will be denoted 
by $\chi ^{\pm}(r;z)$ and 
$\chi ^{\pm}(r;q)$. In bra-ket notation, the analytically
continued eigenfunctions will be written as
\begin{eqnarray}
       \langle r|q^{\pm}\rangle =\chi ^{\pm}(r;q) \, ,  \\
       \langle ^{\pm}q|r \rangle =\chi ^{\mp}(r;q) \, .
\end{eqnarray}
Note that, in distinction to~(\ref{pmeigndubra}), the analytically continued
``left'' eigenfunction is not the complex conjugate of the
``right'' eigenfunction but
\begin{equation}
  \langle ^{\pm}q|r \rangle =\overline{\langle r|\overline{q}^{\pm}\rangle}
      \, .
      \label{lrconus}
\end{equation}

In order to characterize the analytic properties of 
$\chi ^{\pm}(r;q)$, it is useful to define
\begin{equation}
     Z_{\pm} \equiv \{ q \in {\mathbb C} \, | \ {\cal J}_{\pm}(q) = 0 \} \, .
\end{equation}
The elements of $Z_+$ are simply the resonance energies. Since 
$\chi (r;q)$ and ${\cal J}_{\pm}(q)$ are analytic in the whole $k$-plane, 
$\chi ^{\pm}(r;q)$ is analytic in the whole $k$-plane except at $Z_{\pm}$, 
where its poles are located.

In order to define the analytically continued Lippmann-Schwinger bras and
kets, we need to know how $\chi ^{\pm}(r;q)$ grow with $q$. To find out, let 
us recall first that the growth of $\chi (r;q)$ is bounded by~\cite{TAYLOR} 
\begin{equation}
      \left| \chi (r;q)\right| \leq C \, 
        \frac{\left|q\right|r}{1+\left|q\right|r} \,  
      e^{|{\rm Im}(q)|r} \, , \quad q\in {\mathbb C} \, .   
      \label{boundrs}
\end{equation}
From Eqs.~(\ref{defiphi+-}) and 
(\ref{boundrs}), it follows that the eigenfunctions $\chi ^{\pm}(r;q)$ 
satisfy
\begin{equation}
    \hskip-1cm  \left| \chi ^{\pm}(r;q) \right| \leq 
         \frac{C}{|{\cal J}_{\pm}(q)|} \, 
      \frac{|q|r}{1+|q|r} \,
      e^{|{\rm Im}(q)|r }  \, . 
     \label{estimateofphi}
\end{equation}
When $q \in Z_{\pm}$, $\chi ^{\pm}(r;q)$ blows up to infinity.

We can further refine the estimates~(\ref{estimateofphi}) by
characterizing the growth of $1/|{\cal J}_{\pm}(q)|$ in different regions
of the complex plane. In the upper half of the complex $k$-plane, the 
inverse of ${\cal J}_{+}(q)$ is bounded:
\begin{equation}
    \frac{1}{\left| {\cal J}_{+}(q) \right|} \leq C \, ,
    \quad    {\rm Im}(q)\geq 0   \, .
    \label{boundinjslp+}
\end{equation} 
In the lower half-plane, $\frac{1}{{\cal J}_{+}(q)}$ is infinite whenever 
$q \in Z_+$. As $|q|$ tends to $\infty$ in the lower half plane, 
we have
\begin{equation}
    \frac{1}{{\cal J}_{+}(q)} \approx 
    \frac{1}{1 - C q^{-2} e ^{2i q b}}  \, , \quad
       (|q|\to \infty \, , \  {\rm Im}(q)<0)   \, .
    \label{boundinjsup+}
\end{equation} 
The above estimates are satisfied by ${\cal J}_{-}(q)$ when we exchange the 
upper for the lower half plane, and $Z_+$ for $Z_-$.

\subsection{The analytic continuation of the Lippmann-Schwinger bras and kets}

The analytic continuation of the Lippmann-Schwinger bras
is defined for any complex wave number $q$ in the
distributional way~(\ref{afGtff1}):
\begin{equation}
    \langle ^{\pm}q|\varphi ^{\pm}\rangle \equiv
       \int_0^{\infty}d r\, \varphi ^{\pm} (r) \chi ^{\mp}(r;q) =
       \int_0^{\infty}d r\,  \langle ^{\pm}q|r\rangle 
        \langle r|\varphi ^{\pm}\rangle \, ,
       \label{LSdefinitionbra+-q}
\end{equation}
where the functions $\varphi ^{\pm}(r)$ belong to a space of test functions
${\Phi}_{\rm exp}$ that will be constructed below. Similarly to the bras, 
the analytic continuation of the Lippmann-Schwinger kets is defined by way 
of~(\ref{afFtff1}):
\begin{equation}
     \langle \varphi ^{\pm}|q^{\pm}\rangle \equiv
       \int_0^{\infty}d r\, 
              \overline{\varphi ^{\pm} (r)} \chi ^{\pm}(r;q) =
     \int_0^{\infty}d r\, \langle \varphi ^{\pm}|r\rangle 
        \langle r|q^{\pm}\rangle \, .
            \label{LSdefinitionket+-q}
\end{equation}
Note that definition~(\ref{LSdefinitionbra+-q}) is actually a slight 
generalization of~(\ref{afGtff1}), due to~(\ref{lrconus}).

The bras~(\ref{LSdefinitionbra+-q}) and kets~(\ref{LSdefinitionket+-q}) are 
defined for all complex $q$ except at those $q$ at which the 
corresponding eigenfunction has a pole. At those poles, one can still define 
bras and kets if in definitions~(\ref{LSdefinitionbra+-q}) and 
(\ref{LSdefinitionket+-q}) one substitutes the eigenfunctions 
$\chi ^{\pm}(r;q)$ by their residues at the pole. 

From the analytic continuation of the bras and kets into any
complex wave number, one can now obtain the analytic continuation of the 
bras and kets into any complex energy of the Riemann surface
(compare with Eqs.~(\ref{brask}) and (\ref{ketsk})):
\begin{equation}
    |z^{\pm}\rangle =\sqrt{\frac{2m}{\hbar ^2}\frac{1}{2q}}\, |q^{\pm}\rangle
    \, ,  \quad  
    \langle ^{\pm}z|=\sqrt{\frac{2m}{\hbar ^2}\frac{1}{2q}}\, \langle ^{\pm}q|
      \, .
\end{equation}

\subsection{Construction of the rigged Hilbert space}

Likewise the bras and kets associated with real energies, the analytic
continuation of the Lippmann-Schwinger bras and kets must be described 
within the rigged Hilbert space rather than just within the Hilbert 
space. We will denote the rigged Hilbert space for the analytically 
continued bras by
\begin{equation}
     {\Phi}_{\rm exp} \subset L^2([0,\infty ),d r) 
      \subset {\Phi}_{\rm exp}^{\prime} \, ,
         \label{rhsexpp}
\end{equation} 
and the one for the analytically continued kets by
\begin{equation}
     {\Phi}_{\rm exp} \subset L^2([0,\infty ),dr) 
      \subset {\Phi}_{\rm exp}^{\times} \, .
     \label{rhsexpt}
\end{equation} 

The functions $\varphi ^{\pm}\in {\Phi}_{\rm exp}$ must satisfy 
the following conditions: 
\begin{eqnarray}
   \hskip-2cm && \bullet \  \mbox{They belong to the maximal invariant 
                 subspace $\cal D$ of} \ H, \ 
                      \mbox{see Eq.~(\ref{misoH}).}
                           \label{condition1exp} \\ [2ex]
   \hskip-2cm &&\bullet \ \mbox{They are such that
   definitions~(\ref{LSdefinitionbra+-q}) and (\ref{LSdefinitionket+-q}) 
make sense.} 
         \label{condition2exp}
\end{eqnarray}
The reason why $\varphi ^{\pm}$ must satisfy condition~(\ref{condition1exp})
is that such condition guarantees that all the powers of the Hamiltonian are 
well defined. Condition~(\ref{condition1exp}), however, is not sufficient to 
obtain well-defined bras and kets associated with complex wave numbers. In 
order for $\langle ^{\pm}q|$ and $|q^{\pm}\rangle$
to be well defined, the wave functions $\varphi ^{\pm}(r)$ must be well 
behaved so the integrals in Eqs.~(\ref{LSdefinitionbra+-q}) and 
(\ref{LSdefinitionket+-q}) converge. Since by 
Eq.~(\ref{estimateofphi}) $\chi ^{\pm}(r;q)$ grow exponentially with $r$, 
the wave functions 
$\varphi ^{\pm}(r)$ have to, essentially, tame real exponentials. If we define
\begin{equation}
     \| \varphi ^{\pm}\|_{n,n'} \equiv \sqrt{\int_{0}^{\infty}d r \, 
    \left| \frac{nr}{1+nr}\, e ^{nr^2/2} (1+H)^{n'}
             \varphi ^{\pm}(r) \right|^2 \, } 
                 \, , \quad n,n'=0,1,2, \ldots \, , 
      \label{normsLSexp}
\end{equation}
then the space ${\Phi}_{\rm exp}$ is given by
\begin{equation}
      {\Phi}_{\rm exp} =
      \left\{ \varphi ^{\pm}\in {\cal D} \, | 
      \ \| \varphi ^{\pm} \|_{n,n'}<\infty \, , \ n,n'=0,1,2,\ldots \right\} .
       \label{phiexp}
\end{equation}
This is just the space of square integrable functions which belong to the
maximal invariant subspace of $H$ and for which the 
quantities~(\ref{normsLSexp})
are finite. In particular, because $\varphi ^{\pm}(r)$ satisfy the 
estimates~(\ref{normsLSexp}), their tails fall off faster than Gaussians. 

From Eq.~(\ref{estimateofphi}), it is clear that the integrals in
Eqs.~(\ref{LSdefinitionbra+-q}) and (\ref{LSdefinitionket+-q})
converge already for functions that fall off at infinity faster than any 
exponential. We have imposed Gaussian falloff because it will allow us to 
perform resonance expansions in Lecture~5.

It is illuminating to compare the space $\Phi$ of Lecture~3 with the space 
${\Phi}_{\rm exp}$ of Eq.~(\ref{phiexp}). Because for real wave numbers the 
Lippmann-Schwinger 
eigenfunctions behave like purely imaginary exponentials, in Lecture~3 we 
only needed to impose on the test functions a polynomial falloff, thereby 
obtaining a space of test functions very similar to the Schwartz space. By 
contrast, for complex wave numbers the Lippmann-Schwinger eigenfunctions blow 
up exponentially, and therefore we need to impose on the test functions an 
exponential falloff that damps such an exponential blowup.

One can now easily show that the kets $|q^{\pm}\rangle$ belong to
${\Phi}_{\rm exp}^{\times}$ and satisfy
\begin{equation}
      H|q^{\pm}\rangle= \frac{\hbar ^2}{2m}q^2 \, |q^{\pm}\rangle \, ,
       \label{keigeeqa} 
\end{equation}
\begin{equation}
      e^{-i Ht/\hbar}|q^{\pm}\rangle =  e^{-i q^2 \hbar t/(2m)}
        |q^{\pm}\rangle  \, .
     \label{timevoketspm1} 
\end{equation}
Similarly, $\langle ^{\pm}q| \in {\Phi}_{\rm exp}^{\prime}$ and
\begin{equation}
      \langle ^{\pm}q|H=\frac{\hbar ^2}{2m} q^2  \langle ^{\pm}q|\, ,
             \label{kpssleftkeofHa}
\end{equation}
\begin{equation}
      \langle ^{\pm}q|e^{-i Ht/\hbar} =  e^{i q^2 \hbar t/(2m)}
       \langle ^{\pm}q| \, .
      \label{timevobraspm2}
\end{equation}

Equations~(\ref{keigeeqa}) and (\ref{kpssleftkeofHa}) can be rewritten in 
terms of the complex energy $z$ as
\begin{eqnarray}
       H|z^{\pm}\rangle=z|z^{\pm}\rangle \, ,
         \label{eigenzi} \\ [1ex]
      \langle ^{\pm}z|H = z \langle ^{\pm}z| \, .
           \label{eigenzibra}
\end{eqnarray} 
Note that~(\ref{eigenzibra}) is not given by
$\langle ^{\pm}z|H = \overline{z}\langle ^{\pm}z|$, as one may naively
expect from formally obtaining~(\ref{eigenzibra}) by Hermitian conjugation
of~(\ref{eigenzi}). 

For a full account of this lecture, the reader can refer to~\cite{LS2}.

\section{Resonance states, and their rigged Hilbert space: Lecture~5}
\label{sec:RHSresonan}

The Gamow states are the state vectors of resonances. They are 
eigenvectors of the Hamiltonian with a complex eigenvalue. The real 
(imaginary) part of the complex eigenvalue is associated with the energy
(width) of the resonance. 

Because self-adjoint operators on a Hilbert space can only have real 
eigenvalues, the Gamow states fit not within the Hilbert space but
within the rigged Hilbert space. In this lecture, we will see that the
Gamow states belong to the rigged Hilbert space of Lecture~4.

Like in Lectures~3 and~4, we will use the spherical shell 
potential~(\ref{sbpotential}) 
and study the zero angular momentum case only. Unlike in Lecture~4, 
we will write most results in terms of the energy, because they 
tend to be simpler than in terms of the wave number. The energy and the wave 
number of a resonance ${\rm R}$ will be denoted by $z_{\rm R}$ and $k_{\rm R}$.

The Gamow eigenfunctions satisfy the Schr\"odinger equation
\begin{equation}
       \left( -\frac{\hbar ^2}{2m}\frac{d ^2}{d r^2}+V(r)\right) 
       u(r;z_{\rm R}) = z_{\rm R} \, u(r;z_{\rm R}) \, ,
	\label{Grse0}
\end{equation}
subject to ``purely outgoing boundary conditions,'' 
\begin{eqnarray}
	&& u(0;z_{\rm R}) = 0 \, , \label{gvlov1}  \\
	&& u(r;z_{\rm R}) \ \mbox{is continuous at} \ r=a,b \, ,  
                        \label{gvlov2}   \\
	&&\frac{d}{d r}u(r;z_{\rm R}) \ 
                     \mbox{is continuous at} 
          \ r=a,b \, ,  \label{gvlov5} \\
        && u(r;z_{\rm R}) \sim e^{i k_{\rm R}r} \ 
                       \mbox{as} \  r\to \infty \, , 
          \label{gvlov6}  
\end{eqnarray}
where~(\ref{gvlov6}) is the ``purely outgoing boundary condition'' 
(POBC). Comparison of (\ref{gvlov1})-(\ref{gvlov6}) with 
(\ref{bcoLS1})-(\ref{bcoLS6}) shows that it is the POBC what 
selects the resonance energies.

For the potential~(\ref{sbpotential}), the only possible
eigenvalues of~(\ref{Grse0}) subject to~(\ref{gvlov1})-(\ref{gvlov6}) are
the zeros of the Jost function,
\begin{equation}
      {\cal J}_+(z_{\rm R})=0 \, .
         \label{ressoncon}
\end{equation}
The solutions of this equation come as a denumerable number of complex 
conjugate pairs $z_n, z_n^*$. The number $z_n=E_n -i \Gamma _n /2$ is the 
$n$th resonance energy, and $z_n^*=E_n +i \Gamma _n /2$ is the 
$n$th anti-resonance energy. The corresponding wave numbers are
\begin{equation}
      k_n=\sqrt{\frac{2m}{\hbar ^2}z_n\,} \, , \quad
      -k_n^*=\sqrt{\frac{2m}{\hbar ^2}z_n^*\,} \, , \quad n=1,2, \ldots \, .
\end{equation}
For the potential~(\ref{sbpotential}), the resonance energies are
simple poles of the $S$ matrix (see~\cite{MONDRADOUBLE} for an example 
of a potential that produces double poles). In order to write expressions 
for resonances and anti-resonances together, we will label the resonances by 
a positive integer $n=1,2,\ldots$ and the anti-resonances by a negative 
integer $n=-1,-2,\ldots$.

The $n$th Gamow eigensolution, $n=\pm 1, \pm 2, \ldots$, reads 
\begin{equation}
      u(r;z_n)=u(r;k_n)= N_n\left\{ \begin{array}{ll}
         \frac{1}{{\mathcal J}_3(k_n)}\sin(k_{n}r)  &0<r<a \\ [1ex]
         \frac{{\mathcal J}_1(k_n)}{{\mathcal J}_3(k_n)}e^{i Q_{n}r}
         +\frac{{\mathcal J}_2(k_n)}{{\mathcal J}_3(k_n)}
                e^{-i Q_{n}r} &a<r<b 
         \\  [1ex]
         e^{i k_{n}r}  &b<r<\infty \, , 
                           \end{array} 
                  \right. 
	\label{dgv0p} 
\end{equation}
where
\begin{equation}
      Q_n=\sqrt{\frac{2m}{\hbar ^2}(z_n-V_0)\,} \, ,
\end{equation}
and $N_n$ is a normalization factor,
\begin{equation}
       N_n^2= i \, \mbox{res} \left[ S(q) \right]_{q=k_n} \, .
\end{equation}
The Gamow eigenfunctions~(\ref{dgv0p}) are related with $\chi ^{\pm}(r;q)$ by
\begin{equation}
      u(r;k_n) = - \frac{\sqrt{2\pi}}{N_n} \,
       {\rm res}\left[ \chi ^+(r;q) \right]_{q=k_n} \, ,
     \label{residukslls}
\end{equation}
\begin{equation}
       u(r;k_n) = i\sqrt{2\pi}N_n \, \chi ^-(r;k_n)  \, .
     \label{valuemkkns}
\end{equation}

Because of~(\ref{valuemkkns}) and (\ref{lrconus}), the ``left'' Gamow 
eigenfunctions read
\begin{equation}
       \langle z_n|r\rangle = [u(r;z_n^*)]^* 
      \, , \quad n=\pm 1, \pm 2, \ldots \, .
               \label{legazn1}
\end{equation}
Thus, the ``left'' Gamow eigenfunction is not 
just the complex conjugate of the ``right'' eigenfunction, but the complex
conjugated eigenfunction evaluated at the complex conjugated energy. Because
the Gamow eigenfunctions satisfy
\begin{equation}
           [u(r;z_n^*)]^* = u(r;z_n) \, , \quad n=\pm 1, \pm 2, \ldots \, ,
               \label{legazn2}
\end{equation}
the ``left'' and the ``right'' Gamow eigenfunctions are actually the same 
eigenfunction,
\begin{equation}
       \langle z_n|r\rangle = [u(r;z_n^*)]^* = u(r;z_n) = 
         \langle r|z_n \rangle \, , \quad n=\pm 1, \pm 2, \ldots \, .
               \label{legazn}
\end{equation}
In terms of the wave number, Eq.~(\ref{legazn}) reads as
\begin{equation}
       \langle k_n|r\rangle = [u(r;-k_n^*)]^* = u(r;k_n) = 
         \langle r|k_n \rangle \, , \quad n=\pm 1, \pm 2, \ldots \, .
               \label{legakn}
\end{equation}

The Gamow eigenfunctions $u(r;z_n)$ are not square integrable and therefore
must be treated as distributions. By treating them as distributions, we will 
be able to generate the Gamow bras and kets. According to~(\ref{afFtff1}), 
the Gamow ket $|z_n\rangle$ associated with the eigenfunction $u(r;z_n)$ 
must be defined as
\begin{equation}
       \langle \varphi |z_n\rangle \equiv
        \int_0^{\infty} dr\, [\varphi (r)]^* \,  u(r;z_n) =
        \int_0^{\infty} dr\, \langle \varphi |r\rangle
           \langle r|z_n\rangle \, , \quad  n=\pm 1, \pm 2, \ldots \, .
             \label{Gketdef}
\end{equation}
Similarly, the Gamow bra associated with the resonance (or 
anti-resonance) energy $z_n$ is defined as
\begin{equation}
       \langle z_n| \varphi \rangle \equiv
        \int_0^{\infty} dr\, \varphi (r) u(r;z_n) =
        \int_0^{\infty} d r\, 
                   \langle z_n|r\rangle \langle r|\varphi \rangle \, , \quad
            n= \pm 1, \pm 2 , \ldots  \, .
             \label{Gbradef}
\end{equation}
From these definitions and from Eqs.~(\ref{residukslls}) and 
(\ref{valuemkkns}), it is clear that
the Gamow bras and kets are accommodated by the rigged Hilbert 
spaces~(\ref{rhsexpp}) and~(\ref{rhsexpt}). 

Within the rigged Hilbert spaces~(\ref{rhsexpp}) and~(\ref{rhsexpt}), it 
holds that the Gamow bras and kets are eigenvectors of the Hamiltonian:
\begin{equation}
      H|z_n\rangle= z_n \, |z_n\rangle \, , \quad n=\pm 1,\pm 2, \ldots  \, ,
       \label{keigeeqaGS}
\end{equation}
\begin{equation}
      \langle z_n|H= z_n \langle z_n| \, , \quad n=\pm 1,\pm 2, \ldots  \, .
             \label{kpssleftkeofHaGS}
\end{equation}
Because their energy is complex, the time evolution of the Gamow bras and 
kets should be time asymmetric. For resonances, it should be that
\begin{equation}
      \langle z_n| e^{-i Ht/\hbar}= 
         e^{i z_n t/\hbar} \langle z_n| \, ,
        \quad {\rm only\ for}\ t<0  \, , \ n=1,2, \ldots \, , 
      \label{adjoiotevbra1ares}
\end{equation}
\begin{equation}
      e^{-i Ht/\hbar}|z_n\rangle =
       e^{-i z_n t/\hbar} |z_n\rangle  \, ,
      \quad {\rm only\ for}\ t>0  \, , \ n=1,2, \ldots \, ,
      \label{adjoiotev2ares}
\end{equation}
whereas for anti-resonances, it should be that
\begin{equation}
      \langle z_n| e^{-i Ht/\hbar}= 
         e^{i z_n t/\hbar} \langle z_n| \, ,
        \quad {\rm only\ for}\ t>0  \, , \ n=-1,-2, \ldots \, , 
      \label{adjoiotevbra1aresta}
\end{equation}
\begin{equation}
      e^{-i Ht/\hbar}|z_n\rangle =
          e^{-i z_n t/\hbar} |z_n\rangle  \, ,
      \quad {\rm only\ for}\ t<0  \, , \ n=-1,-2, \ldots \, .
      \label{adjoiotev2aresta}
\end{equation}

If we define the complex delta function at $z_n$ by
\begin{equation}
       \int_0^{\infty}dE \, f(E) \delta (E-z_n) = f(z_n) \, ,
\end{equation}
one can use the results of Lecture~4 to show that the Gamow eigenfunction 
$u(r;z_n)$, the complex delta function (multiplied by a normalization factor) 
and the Breit-Wigner amplitude (multiplied by a normalization factor) 
are linked with each other:
\begin{equation}
    \begin{array}{ccccc}
       u(r;z_n)& \leftrightarrow  & 
      i \sqrt{2\pi \, } {\cal N}_n \delta (E-z_n) \, , \ E\in [0,\infty ) & 
      \leftrightarrow  & -\frac{{\cal N}_n}{\sqrt{2\pi}}\,
       \frac{1}{E-z_n}\, , \ E\in (-\infty ,\infty)  \\ [2ex]
       \mbox{posit. repr.} &\ &  \mbox{energy repr.} & \ & 
      (-\infty,\infty)\mbox{-``energy'' repr.}
    \end{array}
    \label{greatres}
\end{equation}
where
\begin{equation}
      {\cal N}_n^2 = i \, {\rm res}[S(z)]_{z=z_n} \, .
\end{equation}
Physically, these links mean that the Gamow states yield a decay amplitude
${\cal A}(z_{\rm R}\to E)$ 
given by the complex delta function, and that such decay amplitude can be 
approximated 
by the Breit-Wigner amplitude when we can ignore the lower bound of the
energy, i.e., when the resonance is so far from the threshold that we
can safely assume that the energy runs over the full real 
line:
\begin{equation}
      {\cal A}(z_{n}\to E)= \langle ^-E|z_{n}\rangle = 
         i \sqrt{2\pi} {\cal N}_{n} \delta (E-z_{n}) 
       \simeq -\frac{{\cal N}_{n}}{\sqrt{2\pi}} \, \frac{1}{E-z_{n}} 
            \, .
     \label{mainresult}
\end{equation}
Thus, the almost-Lorentzian peaks in cross sections are caused by 
intermediate, unstable particles. However, because there is actually a lower 
bound for the energy, the decay amplitude is never exactly given by the 
Breit-Wigner amplitude. This means, in particular, that the standard 
Gamow states are different from the so-called ``Gamow vectors'' 
of~\cite{IJTP03}.

\subsection{Resonance expansions}

The scattering bras and kets are basis vectors that furnish the completeness
relation~(\ref{crHS}). The Gamow states are also basis vectors. The 
completeness relation~(\ref{crRHS}) generated by the Gamow states is called 
the resonance expansion. 

Resonance expansions are almost always obtained in the same way. One 
starts from the expansion in terms of bound and scattering 
states and then, by deforming the continuum integral into the complex plane,
and by Cauchy's theorem, one extracts the contributions from the resonances
that are hidden in the continuum and write them in the same way as the
contributions from the bound states. 

For the sake of simplicity, we will focus on the resonance expansion 
of the transition 
amplitude from an ``in'' state $\varphi ^+$ into an ``out'' state 
$\varphi ^-$:
\begin{equation}
        \left( \varphi ^-,\varphi ^+\right)=
        \int_0^{\infty} d E \, \langle \varphi ^-|E^-\rangle S(E)
        \langle ^+E|\varphi ^+\rangle \, .
        \label{superequation}
\end{equation}
We now extract the resonance contributions out of~(\ref{superequation}) by 
deforming the contour of integration into the 
lower half plane of the second sheet of the Riemann surface, where the 
resonance poles are located, and by applying
Cauchy's theorem. Assuming that the integrand 
$\langle \varphi ^-|E^-\rangle S(E) \langle ^+E|\varphi ^+\rangle$ tends to 
zero in the infinite arc of the lower half plane of the second sheet, the
resulting resonance expansion is
\begin{equation}
      \left( \varphi ^-,\varphi ^+\right)
      =  \sum_{n=1}^{\infty}
      \langle \varphi ^-|z_n\rangle\langle z_n|\varphi ^+\rangle +
    \int_0^{-\infty} d E \, \langle \varphi ^-|E^-\rangle S(E) 
      \langle ^+E|\varphi ^+\rangle  \, . 
      \label{zigzag}
\end{equation}
The integral in Eq.~(\ref{zigzag}) is supposed to be done infinitesimally 
below the negative real semiaxis of the second sheet. By omitting
the wave functions in~(\ref{zigzag}), we obtain the completeness 
relation~(\ref{crRHS}). In Eq.~(\ref{zigzag}), the infinite sum contains the 
contribution from the resonances, while the integral is the 
non-resonant background. Note that in~(\ref{zigzag}) bound states do not
appear, since the potential~(\ref{sbpotential}) doesn't bind any.

In obtaining Eq.~(\ref{zigzag}), we have assumed that the integrand 
$\langle \varphi ^-|E^-\rangle S(E) \langle ^+E|\varphi ^+\rangle$ tends to 
zero in the infinite arc of the lower half plane of the second 
sheet. Since~(\ref{zigzag}) makes physical sense, we
are tempted to conclude that such must be the case. However, as showed 
in~\cite{LS2}, 
$\langle \varphi ^-|E^-\rangle S(E) \langle ^+E|\varphi ^+\rangle$ does 
not tend to zero in the infinite arc of the lower half plane of the 
second sheet. On the contrary, it diverges exponentially there. Therefore, 
Eq.~(\ref{zigzag}) doesn't
make sense as it stands. In order to make sense of it, one has to control
the exponential blowup of 
$\langle \varphi ^-|E^-\rangle S(E) \langle ^+E|\varphi ^+\rangle$. The
way to do so is by introducing an exponentially damping regulator
$e^{-i \alpha z}$, $\alpha >0$. Thus, Eq.~(\ref{zigzag}) should read
\begin{equation}
      \left( \varphi ^-,\varphi ^+\right)
      = \lim_{\alpha \to 0}  \sum_{n=1}^{\infty} e^{-i \alpha z_n}
      \langle \varphi ^-|z_n \rangle\langle z_n|\varphi ^+\rangle +
    \int_0^{-\infty} d E \,
       e^{-i \alpha E} \langle \varphi ^-|E^-\rangle S(E) 
      \langle ^+E|\varphi ^+\rangle .
      \label{alphazigzag}
\end{equation}
Physically, the regulator $e^{-i \alpha z}$ is simply the analytic continuation
of the time evolution operator in the energy representation,
$e^{-i z \alpha} \equiv e^{-i zt/\hbar}$. Thus, the above
regularized equation must be understood in a time-asymmetric,
time-dependent fashion as
\begin{equation}
      (\varphi ^-, e^{-i H\frac{t}{\hbar}} \varphi ^+ )
      = \sum_{n=1}^{\infty} e^{-i z_n\frac{t}{\hbar}}
      \langle \varphi ^-|z_n\rangle \langle z_n|\varphi ^+\rangle +
    \int_0^{-\infty} dE \,
       e^{-i E\frac{t}{\hbar}} \langle \varphi ^-|E^-\rangle S(E) 
      \langle ^+E|\varphi ^+\rangle  
            \label{alphazigzagtime}
\end{equation}
for $t>0$ only. Equation~(\ref{zigzag}) should then be seen as the 
(singular) limit of Eq.~(\ref{alphazigzagtime}) when
$t \to 0^+$. That for resonances $t \equiv \alpha  \hbar$ must be positive
is in accord with the time asymmetry of~(\ref{adjoiotev2ares}).

Expansions~(\ref{alphazigzag}) and (\ref{alphazigzagtime}) are the reason 
why we chose a Gaussian falloff for the elements of $\Phi _{\rm exp}$: When 
the wave functions have a Gaussian falloff in the position representation, 
we can regularize their blowup in the energy representation and interpret the 
regulator as a time-asymmetric evolution. 

Resonance expansions allow us to understand the deviations from exponential 
decay. When a particular resonance, say resonance 1, is dominant, then
the Gamow state of resonance~1 will carry the exponential decay, whereas the
background, which includes in this case also the contribution from other
possible resonances, carries the deviations from exponential decay.

The full account of this lecture will appear in a forthcoming paper.

\section{Conclusions}
\label{sec:conclusions}

We have seen why the rigged Hilbert space, rather than the Hilbert space
alone, is needed to formulate quantum mechanics when the observables have 
continuous and/or resonance spectra. The rigged Hilbert space captures the 
physics of continuous and resonance spectra better than
the Hilbert space, because in the rigged Hilbert space physical quantities
such as commutation relations, uncertainty principles and resonances have
always a precise meaning. 

In addition to provide the mathematical support for Dirac's bra-ket formalism, 
for the Lippmann-Schwinger equation and for the Gamow states, the rigged 
Hilbert space can be used to obtain the resonance (decay) amplitude in terms 
of the complex delta function. Such decay amplitude can be 
approximated by the Breit-Wigner amplitude when the lower bound of the 
energy can be ignored.

To finish, I would like to mention that there is still a long list of pending
questions worth pursuing, such as the invariance properties of
$\Phi _{\rm exp}$ under time evolution or a detailed proof of the 
asymmetry in the time evolution of the Gamow states.

\begin{theacknowledgments}

It is a pleasure to thank the organizers for their invitation to participate 
in this summer school and Oscar Rosas-Ortiz for his hospitality. This work was 
supported by MEC fellowship No.~SD2004-0003.

\end{theacknowledgments}


\begin{thebibliography}{99}


\bibitem{PERES} A.~Peres, \emph{Quantum Theory: Concepts and Methods},
Dordrecht, Kluwer Academic (1993).

\bibitem{IJTP03} A.~Bohm, Int.~J.~Theo.~Phys.~{\bf 42}, 2317 (2003).

\bibitem{HARDY} R.~de la Madrid, J.~Phys.~A: Math.~Gen.~{\bf 39}, 9255
(2006); {\sf quant-ph/0606186}.

\bibitem{DIS} R.~de la Madrid, \emph{Quantum Mechanics in Rigged Hilbert
Space Language}, PhD Dissertation, Universidad de Valladolid, Spain 
(2001). Available at \url{http://www.physics.ucsd.edu/~rafa}.

\bibitem{04JPA} R.~de la Madrid, J.~Phys.~A: Math.~Gen.~{\bf 37}, 8129 (2004); 
{\sf quant-ph/0407195}.

\bibitem{05EJP} R.~de la Madrid, Eur.~J.~Phys.~{\bf 26}, 287 (2005);
{\sf quant-ph/0502053}.

\bibitem{LS1} R.~de la Madrid, J.~Phys.~A:~Math.~Gen.~{\bf 39}, 3949 (2006);
{\sf quant-ph/0603176}. 

\bibitem{TAYLOR} J.~R.~Taylor, \emph{Scattering theory}, John Wiley \& 
Sons, Inc., New York (1972).

\bibitem{LS2} R.~de la Madrid, J.~Phys.~A:~Math.~Gen.~{\bf 39}, 3981 (2006);
{\sf quant-ph/0603177}. 

\bibitem{MONDRADOUBLE} E.~Hern\'andez, A.~J\'auregui, A.~Mondrag\'on, 
J.~Phys.~A: Math.~Gen.~{\bf 33}, 4507 (2000).

\end{thebibliography}
\end{document}